\documentclass{ws-p8-50x6-00}

\begin{document}

\title{Bottom Production in Two-Photon Collisions at LEP}

\author{Armin B\"ohrer}

\address{Fachbereich Physik, Universit\"at Siegen, 57068 Siegen, Germany\\
E-mail: armin.boehrer@cern.ch}


\maketitle

\abstracts{Results on inclusive and exclusive bottom production in 
$\gamma\gamma$ collisions are presented. The total cross section of 
inclusive bottom production is investigated through its leptonic 
decays at LEP~II energies by the experiments L3 and OPAL. 
The average cross section, after correction for acceptances and efficiencies, 
is $\sigma_{\mathrm {tot}} = 13.3 \pm 1.5 \pm 2.3 \,{\mathrm {pb}}$. 
The next-to-leading order calculations are lower than the data by a factor 
three, which corresponds to a difference of more than three 
standard deviations.\\
ALEPH studied the exclusive bottom production. 
Searching for the $\eta_{\mathrm b}$ one candidate is found. Limits on 
$\Gamma_{\gamma\gamma}(\eta_{\mathrm b}) \times$BR for 4 and 6 charged 
particles are extracted. The candidate has a mass of 
$9.30 \pm 0.04\,{\mathrm{GeV}}$.}

\section{Introduction}

\begin{picture}(10,0)(0,0)
\put(370.,224.){SI-2001-5}
\put(370.,211.){May 2001}
\end{picture}
The production of heavy flavour in two-photon collisions is dominated
by two processes, the direct and the single-resolved process. Both contribute
in equal shares to heavy flavour final states at
LEP~II energies. The large quark
mass allows reliable perturbative calculations for the direct contribution.
The single-resolved one, in addition, depends on the gluon density of the
photon.

In this article the new measurements on inclusive\cite{allincl} and 
exclusive\cite{allexcl} bottom production at LEP will be presented. 
The charm production in $\gamma\gamma$ collisions is discussed in 
a separate contribution\cite{talkcharm}. 

\section{Inclusive Bottom Production}

\subsection{Analysis}

Open bottom production is measured by the L3 and the OPAL 
collaborations\cite{allincl} at LEP~II energies using an integrated 
luminosity of 
$400\,{\mathrm{pb^{-1}}}$. Their analysis procedures exploit the
fact, that the momentum as well as the transverse momentum of leptons
with respect to the closest jet is higher for muons and electrons from bottom
than from background, which is mainly charm. Therefore, leptons with momenta 
of more than $2\,{\mathrm{GeV}}$ are selected and their momentum distribution 
with respect to the closest jet (obtained with the JADE jet-algorithm in L3 
and KTCLUS in OPAL, while in both experiments the lepton was excluded, 
when defining the jet) is investigated. 

\begin{figure}

\begin{minipage}{.48\textwidth}

\vspace*{-0.83cm}

\includegraphics[width=1.00\textwidth,clip]{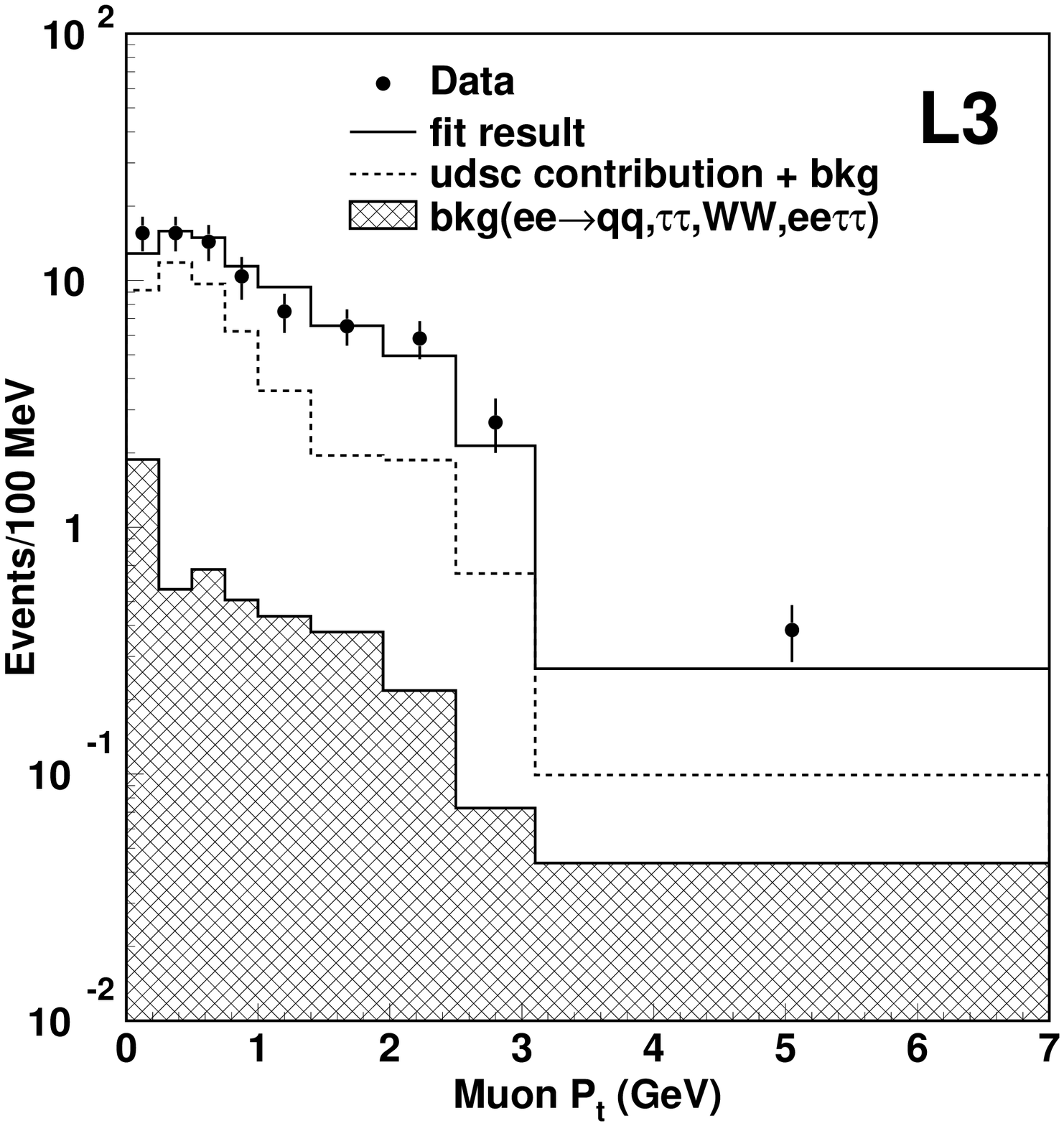}
\caption{Transverse momentum of electrons after fit of the signal 
and background contributions}
\label{bottomL3}
\end{minipage}
\begin{minipage}{.04\textwidth}
~
\end{minipage}
\begin{minipage}{.48\textwidth}
\includegraphics[width=1.00\textwidth,clip]
{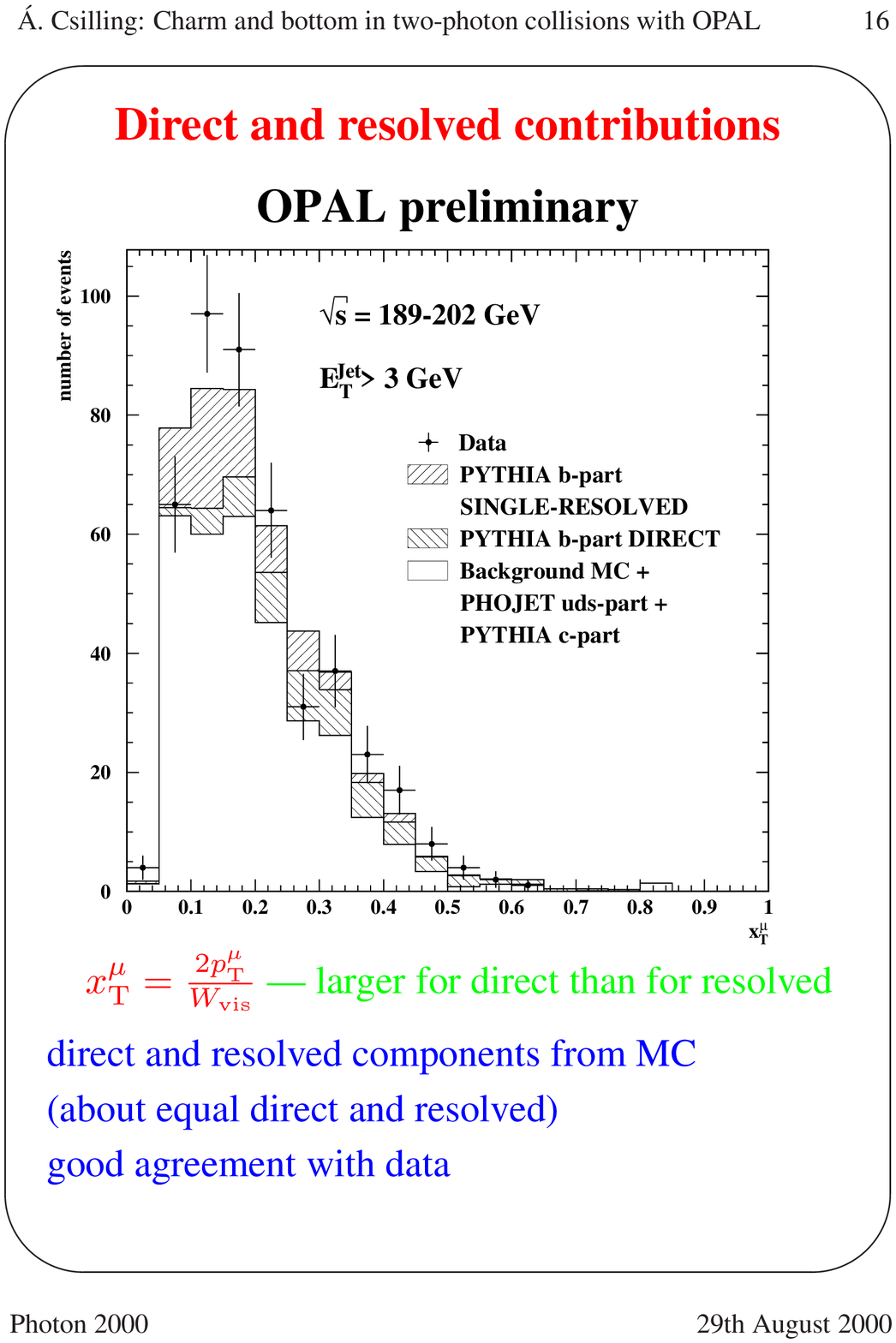}
\caption{direct and single resolved contributions from the 
$x_{\mathrm T}^{\mu} = 2p_{\mathrm T}^{\mu}/W_{\mathrm{vis}}$ distribution}
\label{bottomOPAL}
\end{minipage}
\end{figure}

The various contributions of the spectrum from signal 
and background are fitted, see Figure~\ref{bottomL3} as an example.
In L3 both muons and electrons 
are selected with several percent efficiency, the b-fraction being 52\% and 
42\%, respectively. OPAL selects their muons with a similar efficiency. 
The b-fraction is determined to 27\%. 

Similar to the studies in charm production, the bottom quarks produced 
in direct and single resolved events show a different behaviour in the 
transverse momentum distribution. The variable $x_{\mathrm T}^{\mu} = 
2p_{\mathrm T}^{\mu}/W_{\mathrm{vis}}$ is well suited to demonstrate the 
need for both contributions: the single resolved part at low 
$x_{\mathrm T}^{\mu}$ and the direct part at high $x_{\mathrm T}^{\mu}$ as 
shown in Figure~\ref{bottomOPAL}. The agreement between data and Monte Carlo 
simulation is very good.

\subsection{Cross Section Results}

The total cross section measurements for open bottom production 
are summarized in Figure~\ref{bottomx}. The results are compared to 
NLO calculations\cite{nlo}. The cross sections as measured by the 
experiments are $13.1 \pm 2.0 \pm 2.4\,{\mathrm {pb}}$ (L3) and 
$14.2 \pm 2.5^{+5.3}_{-4.8}\,{\mathrm {pb}}$ (OPAL). The calculations 
underestimate the data by a factor 3 corresponding to $2.5$ and $2\sigma$ 
standard deviations, respectively.

\section{Exclusive Bottom Production}

\subsection{Motivation}

The ALEPH experiment has started a search for the still undiscovered 
$\eta_{\mathrm b}$ pseudoscalar meson. Various predictions exist for 
the mass of the $\eta_{\mathrm b}$, e.g., from potential models, pQCD, 
NRQCD, and lattice calculations. While the production can reliably be 
estimated (about 156 $\eta_{\mathrm b}$ mesons were produced in ALEPH for 
an integrated luminosity of $700\,{\mathrm{pb^{-1}}}$) the branching ratios 
of the meson have to be guessed. The efficiencies for the two decay modes 
of the $\eta_{\mathrm b}$ under study (4 charged particles or 6 charged 
particles) are around 16\% and 10\%. For branching ratios between 2\% and 5\% 
one would expect 0.5 to 1 event in each channel. The background is 
estimated to be of the same order.

\subsection{Results}

\begin{figure}
\begin{minipage}{.48\textwidth}
\includegraphics[width=1.00\textwidth,clip]{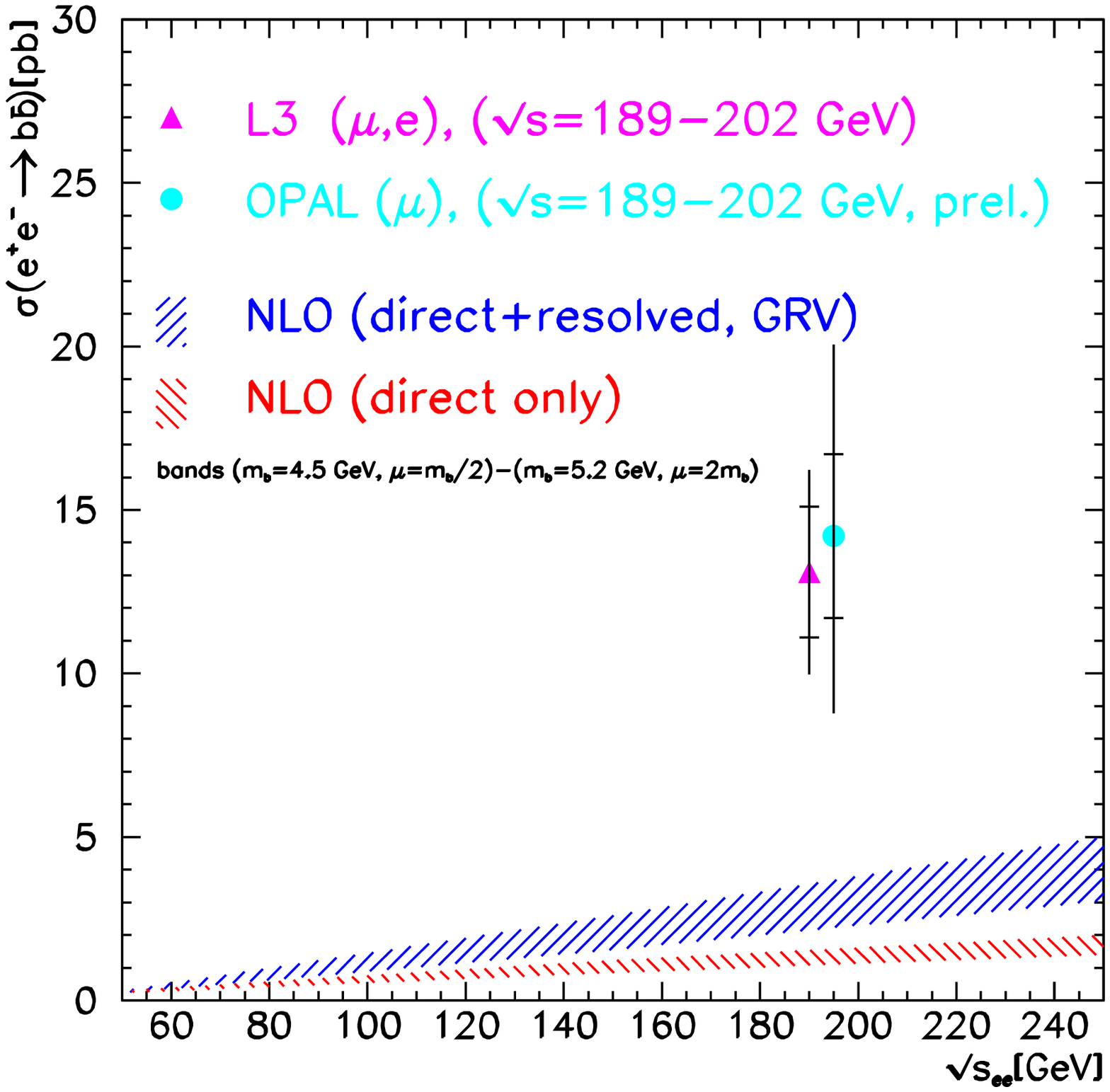}
\caption{Measured bottom cross section in comparison to NLO-calculations. The 
data are shown with statistical and total error. The theoretical bands include 
uncertainties of the b-mass and the scale $\mu$}
\label{bottomx}
\end{minipage}
\begin{minipage}{.04\textwidth}
~
\end{minipage}
\begin{minipage}{.48\textwidth}
\includegraphics[width=1.00\textwidth,clip]{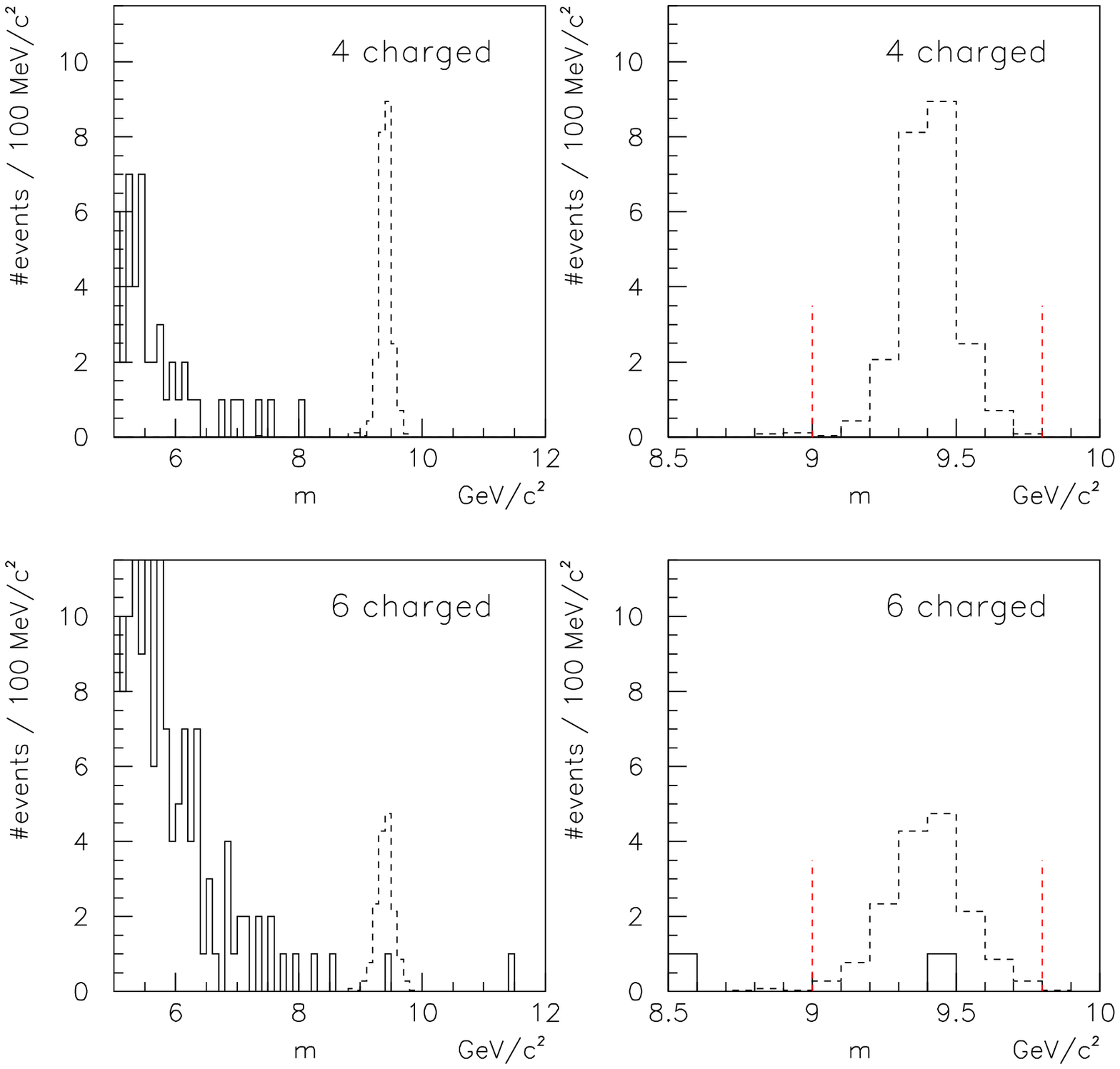}
\caption{Invariant mass distributions in the search for the $\eta_{\mathrm b}$ 
meson. Data (solid line) are compared to the expected signal for an assumed 
branching ratio of 100\% for the decay to 4 and 6 charged particles}
\label{etabinvmass}
\end{minipage}
\end{figure}

In Figure~\ref{etabinvmass} the invariant mass spectra for data are shown 
together with the expected signal assuming branching ratios of 100\% for 
the decay modes under study. No event is selected in the signal region 
from $9.0\,{\mathrm{GeV}}$ to $9.8\,{\mathrm{GeV}}$ in the 4 charged 
mode, while 1 candidate is selected in the 6 charged mode. After 
proper mass assignment the mass of the candidate is 
$9.30 \pm 0.04\,{\mathrm{GeV}}$.

The observation of 0 and 1 candidate, being compatible with background, is 
converted into limits at 95\%CL: 
$\Gamma_{\gamma\gamma}(\eta_{\mathrm b}) \times$BR(4cha)$<57\,{\mathrm{eV}}$ 
and 
$\Gamma_{\gamma\gamma}(\eta_{\mathrm b}) \times$BR(6cha)$<128\,{\mathrm{eV}}$, 
corresponding to 
BR($\eta_{\mathrm b} \rightarrow 4 \, {\mathrm{charged}}) < 17\%$ and 
BR($\eta_{\mathrm b} \rightarrow 6 \, {\mathrm{charged}}) < 38\%$.

\section{Summary}

The LEP experiments have provided good measurements of the inclusive 
bottom production in $\gamma\gamma$ collisions. The combined result 
of the L3 and OPAL experiment is $\sigma_{\mathrm{tot}} 
= 13.3 \pm 1.5 \pm 2.3 \,{\mathrm {pb}}$. The NLO-calculations 
underestimate the cross section by a factor three: they differ by 
more than three standard deviations.

The $\eta_{\mathrm b}$ meson has been searched for. Both expected signal and 
background are about one event. With one candidate observed limits are given 
by the ALEPH experiment: 
$\Gamma_{\gamma\gamma}(\eta_{\mathrm b}) \times$BR(4cha)$<57\,{\mathrm{eV}}$ 
and 
$\Gamma_{\gamma\gamma}(\eta_{\mathrm b}) \times$BR(6cha)$<128\,{\mathrm{eV}}$. 
A discovery would need the effort of all 4 experiments.

\end{document}